# Experimental realization of Controlled Square Root of Z Gate Using IBM's Cloud Quantum Experience Platform

Experimentelle Realisierung der kontrollierten Quadratwurzel des Z-Gatters mit der Cloud-Quantum-Erfahrungsplattform von IBM

Petar N. Nikolov[*], Vassil T. Galabov[†]

[*]FDIBA, Technical University - Sofia
Sofia, Bulgaria, petar.nikolov@fdiba.tu-sofia.bg, vtg@tu-sofia.bg

*Abstract* — Quantum computers form a technological cluster with huge growth in the last few years. Although this technology is of still very limited size – perhaps the reason it is not seen as a technology which may be mass produced or of public use in the near future– it is one of the most promising developments with a potential to change the world. The IBM Quantum Experience Platform makes it possible for every person around the world, without limitation as to geographical location, to get acquainted with the technology of quantum computing. It is a resource for both researchers and enthusiasts entering the quantum world. With the development of the platform, IBM has proven that the programming and writing code executable on a quantum computer can be easy and accessible (it is a cloud platform) even to people lacking any deep knowledge of quantum mechanics.

The construction of the Controlled Square Root of Z gate is achieved using only existing predefined gates in the Composer tool. This newly created gate could be used for further work on quantum algorithms and opens a new feasible way to write quantum code.

*Zusammenfassung* — Quantencomputer bilden in den letzten Jahren einen technologischen Cluster mit einem enormen Wachstum. Obwohl diese Technologie immer noch sehr begrenzt ist - vielleicht der Grund, warum sie nicht als Technologie betrachtet wird, die in naher Zukunft massenhaft oder öffentlich genutzt werden kann - ist sie eine der vielversprechendsten Entwicklungen mit einem Potenzial, die Welt zu verändern. Die IBM Quantum Experience Platform ermöglicht es jedem Menschen auf der ganzen Welt, ohne Einschränkung der geographischen Lage, die Technologie des Quantencomputers kennenzulernen. Es ist eine Ressource für Forscher und Enthusiasten, die in die Quantenwelt eindringen. Mit der Entwicklung der Plattform hat IBM bewiesen, dass der Programmier- und Schreibcode, der auf einem Quantencomputer ausführbar ist, einfach und zugänglich sein kann (es ist eine Cloud-Plattform), auch für Menschen, die keine tiefe Kenntnis der Quantenmechanik haben.

Die Konstruktion der kontrollierten Quadratwurzel des Z-Gates wird unter Verwendung nur vorhandener vordefinierter Gatter im Composer-Werkzeug erreicht. Dieses neu geschaffene Gate könnte für weitere Arbeiten an Quantenalgorithmen verwendet werden und eröffnet eine neue Möglichkeit, Quantencode zu schreiben.

## I. Introduction

A qubit is the equivalent of a bit in classical computing. It takes a value of 0 or 1 when measured [4]. Unmeasured, however, the qubit is in superposition $|\psi\rangle$, which gives the mathematical representation of the qubit at any given time as a two-dimensional state space with orthonormal basis vectors $|0\rangle$ and $|1\rangle$:

$$|\psi\rangle = a_0|0\rangle + a_1|1\rangle \quad (1)$$

The quantum probabilities $a_0$ and $a_1$ represent the chance that a given quantum state will be observed when the superposition has collapsed. These probabilities are complex numbers that satisfy the conditions:

$$|a_0^2| + |a_1^2| = 1 \quad (2)$$

$$\||\psi\rangle\| = \langle\psi|\psi\rangle = 1 \quad (3)$$

When the qubits are connected into strings, the result is a quantum register. The length of the string determines the amount of information this register can store. The superposition of a register $|\psi_n\rangle$ means that each qubit in this register is in superposition.

$$|\psi_n\rangle = \sum a_i|i\rangle \quad (4)$$

So a register of n qubits is in superposition of all $2^n$ possible bit strings, that could be represented using n bits (4), where i is a bit string of 0s and 1s.

The operations over quantum registers are performed by Quantum logic gates [3]. Quantum logic gate applied to a quantum registers maps one quantum superposition to another, together allowing the evolution of the system to some desired final state.

The mathematical representation of quantum logic gates involves transformation matrices, or linear operators, applied to a quantum register by tensoring the transformation matrix with the matrix representation of the register. All linear operators that correspond to quantum logic gates must be unitary ( if a matrix U is unitary, U⁻¹ is equal to the conjugate transpose Û⊤):

$$U^{-1} = \hat{U}^T \quad (5)$$

## II. QUANTUM CONTROLLED SQUARE ROOT OF Z GATE

Quantum computing uses controlled operations – multi-qubit operations that change the state of a qubit based on the values of other qubits [5].

Knowing that the Z gate has the following matrix representation:

$$Z = \begin{bmatrix} 1 & 0 \\ 0 & -1 \end{bmatrix} \quad (6)$$

and given that Z = √Z√Z, a conclusion can be made that the √Z has the following matrix representation:

$$\sqrt{Z} = \begin{bmatrix} 1 & 0 \\ 0 & i \end{bmatrix} \quad (7)$$

where i is the imaginary unit ( $i^2 = -1$).

The controlled-√Z gate is a gate that operates on two qubits in such a way that the first qubit serves as a control qubit for the second one. It has the matrix representation

$$C\sqrt{Z} = \begin{bmatrix} 1 & 0 & 0 & 0 \\ 0 & 1 & 0 & 0 \\ 0 & 0 & 1 & 0 \\ 0 & 0 & 0 & i \end{bmatrix} \quad (8)$$

## III. EXPERIMENTAL RESULTS FROM THE IBM'S QUANTUM EXPERIENCE PLATFORM

IBM's Quantum Experience Platform is a 5-qubit quantum computer available on the cloud. In this platform writing a quantum code can be achieved by a tool called Compozer. In this tool we have a few limitations, one of which is the available gates we can use for computation. The gates we are going to use for the experimental realization of the controlled-√Z gate ( Fig. 2, Fig. 3 and Fig. 5) are as follows:

- Identity Gate – It performs an idle operation on the qubit.

- Pauli-X Gate – It performs a π-rotation around the X axis on a qubit. (Bit flip)

- Controlled NOT Gate – It flips the target qubit if the control is in state 1.

- Phase gate that is square root of S gate, where S gate is √Z gate.

- Phase gate that is transposed conjugate of square root of S gate

- Measurement in the standard basis (Z)

Adding a control qubit to a √Z gate creates a simple quantum circuit. It applies a √Z gate to the target qubit if the control qubit is in state |1> . In general we follow the rules that:
HXH = Z
SXS⊤ = Y
HH = I
SS⊤ = I

Where H is Hadamard Gate, X is Pauli-X Gate, Z is Pauli-Z Gate, S is Phase Gate that is √Z gate, S⊤ is the conjugate transpose of S Gate, Y is Pauli-Y Gate.

Every quantum operation is a unitary matrix and every unitary matrix can be realized as a quantum operation [2]. Every unitary matrix has a square root, so every quantum operation also has a square root. It is possible to factor a controlled quantum operation into pieces [1], so we will be able to find matrices A,B and C and a phase vector $e^{i\Theta}$ such that [5]:

$$e^{i\theta}A \cdot X \cdot B \cdot X \cdot C = U \quad (9)$$

and

$$A.B.C = I \quad (10)$$

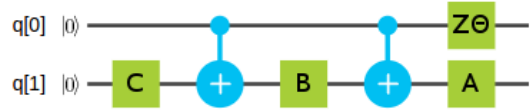

Fig. 1. Controlled U Gate

In Fig. 1 we see that the measured output will be the same as the input when the q[0] input state is |0> because of (10). And the ZΘ does not fire, because phase-shift gates act only on |1> state. In our case the appropriate values for the A, B, C and Θ are as follows [5, 6]:

- We can skip the C
- B = ⁴√Z⊤
- A = ⁴√Z
- ZΘ = ⁴√Z

In IBM's Quantum Experience Platform the √Z gate is represented by the S gate. The ⁴√Z is represented by the T gate and its conjugate transpose is represented by the T⊤ gate. The final result for the Controlled Square root of Z gate is shown in the Fig. 2.

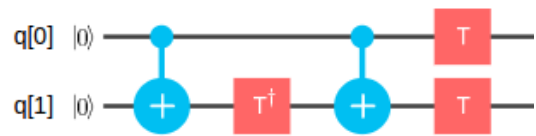

Fig. 2. Controlled Square Root of Z Gate

Because the constructed gate acts on two qubits, four different input combinations could be possible, depending on the qubit states before applying the gate. When the initial states of the two qubits are |11> a phase shift for the q[1] with π/2 radians is observed (Fig. 3 and Fig. 4).

The Quantum Sphere in Fig. 4 and Fig. 6 represents geometrically the pure state space of qubit in Hilbert space, where the north and the south pole correspond to the standard basis vectors |0> and |1>.

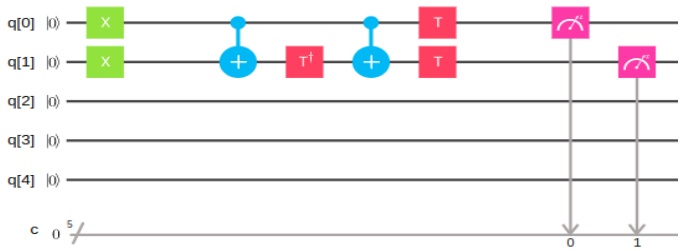

Fig. 3. The IBM Q – Controlled Square Root of Z Gate Circuit |11>

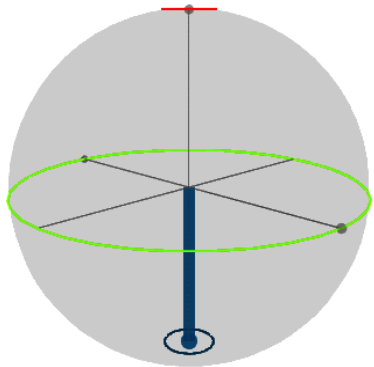

Fig. 4. The IBM Q – Quantum State: Quantum Sphere |11>

When the initial states of the two qubits are |01> we do not have phase shift for the q[1] because the control qubit does not fire (Fig. 5 and Fig. 6).

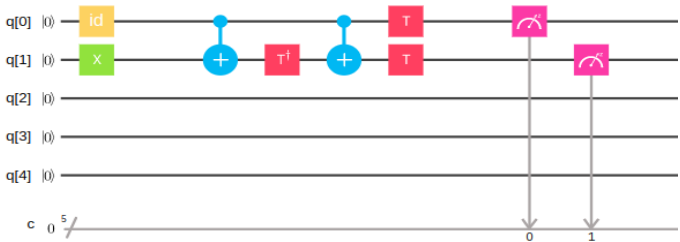

Fig. 5. The IBM Q – Controlled Square Root of Z Gate Circuit |01>

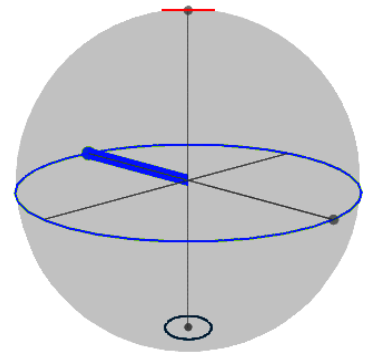

Fig. 6. The IBM Q – Quantum State: Quantum Sphere |01>

## Conclusion

Our experiment for the realization of Controlled Square Root of Z Gate on IBM's Quantum Experience Platform was presented in this paper. Ideally, this work will provide a workable solution for the construction of other quantum gates on this platform in the future, which might contribute for the research and development of new quantum algorithms.

## Acknowledgment

We acknowledge the use of IBM's Quantum Experience Platform for this work. All the views expressed are those of the authors and do not reflect the official position or policy of IBM. We are very grateful to the group for innovative bioinformatics research at Sofia Tech Park BioInfoTech Bioinformatics Lab for their support.